\documentclass[12pt]{article}

\usepackage{amsfonts,epsfig,graphicx}
\usepackage{amsmath,amssymb,amsthm}
\usepackage{epsf}
\usepackage{graphics}
\usepackage{amsfonts}
\usepackage{amsmath}
\usepackage{amssymb}
\usepackage[usenames,dvipsnames]{color}
\usepackage{graphicx}
\usepackage{times}

\usepackage{latexsym}

\usepackage[colorlinks,citecolor=blue]{hyperref}
\usepackage{tikz}
\usetikzlibrary{calc,positioning}
\usepackage{bbm}
\usepackage{float}
\usepackage{enumitem}
\usepackage{comment}

\usepackage{MnSymbol}

\usepackage[left=2cm, right=2cm, bottom=2cm]{geometry}
\newcommand{\mylabel}{\label}
\newcommand{\mybibitem}{\bibitem}

\newcommand{\mB}{\mathcal{B}}
\newcommand{\mF}{\mathcal{F}}
\newcommand{\mM}{\mathcal{M}}
\newcommand{\mP}{\mathcal{P}}

\newcommand{\mbR}{\mathbb{R}}

\newtheorem{thm}{Theorem}
\newtheorem{prop}{Proposition}
\newtheorem{lem}{Lemma}
\newtheorem{cor}{Corollary}

\newtheorem{rem}{Remark}

\begin{document}
\begin{center}
\textbf{\Large A VARIATIONAL CHARACTERIZATION OF R\'{E}NYI DIVERGENCES}
\end{center}

\vspace{.5in}

\begin{center}
VENKAT ANANTHARAM\footnote{EECS Department, University of California, Berkeley, CA 94720, USA. Research supported in part by the
National Science Foundation grants ECCS-1343398, CNS-1527846, 
CCF-1618145, the NSF Science
\& Technology Center grant CCF-0939370 (Science of Information), and the 
William and Flora Hewlett Foundation Center for Long Term Cybersecurity at Berkeley.} 
\end{center}

\vspace{1in}

\noindent \textbf{\large ABSTRACT:} 
Atar, Chowdhary and Dupuis
have recently exhibited a variational formula for
exponential integrals of bounded measurable functions in terms of R\'{e}nyi divergences.
We develop a variational characterization of the R\'{e}nyi divergences between two probability 
distributions on a measurable space in terms of relative entropies. When
combined
with the elementary variational formula for exponential integrals of bounded measurable
functions in terms of relative entropy, this
yields the variational formula of Atar, Chowdhary and Dupuis 
as a corollary.

We also develop an analogous variational characterization of the R\'{e}nyi divergence
rates between two stationary finite state Markov chains in terms of relative entropy rates.
When combined with Varadhan's variational characterization of the spectral radius of
square matrices with nonnegative entries in terms of relative entropy,
this yields an analog of the 
variational formula of Atar, Chowdary and Dupuis in the framework of finite state
Markov chains.
\\


\newpage

\section{Introduction}		\mylabel{sec:Intro}

Evaluating how far away a given probability distribution is from another can be done in
many ways. The Kullback-Leibler divergence or relative entropy, which is closely tied to 
Shannon's notion of entropy, is one such measure prominent in statistical applications. 
It belongs to a larger
family of divergences, the so-called R\'{e}nyi divergences, which are closely tied to 
R\'{e}nyi's notion of entropy. R\'{e}nyi divergences also have numerous applications in 
problems of interest in statistics and information theory, see \cite{EH} for a survey
of some of their basic properties and some indication of their applications.
The R\'{e}nyi divergences, with a minor change in scaling relative to the definition in 
\cite{EH}, are the topic of this article. We treat the R\'{e}nyi divergences as parametrized
by a real number $\alpha \in \mbR$, $\alpha \neq 0$, $\alpha \neq 1$. 

We were prompted to write this document by reading a recent paper of Atar, Chowdhary and
Dupuis \cite{ACD}, which provides a variational formula for
exponential integrals of bounded measurable functions in terms of R\'{e}nyi divergences.
We show that the variational characterization in \cite{ACD}
is a simple consequence of a variational characterization for R\'{e}nyi divergences in terms
of relative entropies, which we also develop. 
For the case of probability distributions on a finite set, and in the range $\alpha > 0$, 
$\alpha \neq 1$, our variational characterization for R\'{e}nyi divergences 
was developed by Shayevitz, \cite{Sha-ISIT} and \cite[Thm. 1]{Sha-Arxiv}.
More recently, for mutually absolutely continuous probability distributions on a 
measurable space, in the case $\alpha > 0$, $\alpha \neq 1$, 
parts of this variational characterization appear
in a paper of Sason, see \cite[Lem. 4 and Cor. 2]{Sason}. 
The ability to derive the variational formula of \cite{ACD} from inequalities for the 
R\'{e}nyi divergences in terms of relative entropies, in the case $\alpha > 1$, is also
remarked on in a recent paper of Liu, Courtade, Cuff, and Verd\'{u} 
\cite[Sec. II-A]{LCCV}.
To the best of our knowledge, however, a full treatment
of this variational characterization of R\'{e}nyi divergences in terms of relative entropies,
covering an arbitrary pair of probability distributions on a measurable space and 
all possible values for $\alpha$,
does not appear to be in the literature and so it seems worth writing down. 
It is also worth noting how easily the full variational formula
of \cite{ACD}, in all cases, falls out of this variational characterization of R\'{e}nyi divergences.

Section \ref{sec:Setup} presents the notational conventions and the definitions of the main 
quantities used in this document in the i.i.d.\ case. The main result in the i.i.d.\ case, Theorem \ref{thm:Main}, is stated in 
Section \ref{sec:Statement}.
The result of \cite{ACD} that prompted this paper is presented in Section \ref{sec:Discussion},
and is derived there as a consequence of Theorem \ref{thm:Main} and the elementary
variational formula for exponential integrals in \eqref{eq:EasyVar}. Theorem \ref{thm:Main}
itself is proved in Section \ref{sec:Proof}.


We then turn to a development of analogs of the preceding results in the case of 
stationary finite state Markov chains. Section \ref{sec:Markov} 
makes the necessary definitions and gathers some standard
facts about the asymptotic properties of iterated powers of a square matrix with nonnegative
entries, which we need for our discussion. It also contains the analog of the elementary 
variational formula in the context of finite state Markov chains, in \eqref{eq:MarkovEasyVar},
which is Varadhan's variational characterization 
in terms of relative entropy of the spectral radius of square matrices with 
nonnegative entries.
The main results in the case of stationary finite state Markov chains are stated in
Section \ref{sec:MarkovMain}. These are Theorem \ref{thm:MarkovMain}, which
gives a variational characterization of each R\'{e}nyi divergence rate between two stationary 
finite state Markov chains in terms of relative entropy rates, and
Theorem \ref{thm:MarkovACD}, which gives an analog of the variational formula of 
\cite{ACD} in the context of finite state Markov chains. A proof of Theorem 
\ref{thm:MarkovACD} assuming the truth of Theorem \ref{thm:MarkovMain}, and using
\eqref{eq:MarkovEasyVar}, is also provided in this section. 
The proof of Theorem \ref{thm:MarkovMain} is provided in Section \ref{sec:MarkovProofMain}.
We end the paper in Section \ref{sec:Conclusions}
with some thoughts about directions for future work. 

In order to maintain the flow of the main exposition, the details of several proofs are relegated to appendices.

\section{Setup}		\mylabel{sec:Setup}

Let $(S,\mF)$ be a measurable space. $\mB(S)$ denotes the set of 
bounded measurable real-valued functions
and $\mP(S)$ the set of probability measures on $(S,\mF)$. 
For $\nu, \theta \in \mP(S)$, 
$\nu \preceq \theta$ is notation for $\nu$ being absolutely 
continuous with respect to $\theta$, see \cite[pg. 442]{Bill} for the definition.
If $\nu \preceq \theta$, then $\frac{d \nu}{d \theta}$ denotes the 
Radon-Nikodym derivative of $\nu$ with respect to $\theta$; any 
two choices of Radon-Nikodym derivative differ only on a $\theta$-null
set, see \cite[Thm. 32.2]{Bill}.
The relative entropy $D(\nu \| \theta)$
of $\nu$ with respect to $\theta$ is defined by
\begin{equation}		\mylabel{eq:RelEnt}
D(\nu \| \theta) := \begin{cases} 
\int_S \left( \log \frac{ d \nu}{d \theta} \right) d \nu~, & \mbox{ if $\nu \preceq \theta$,}\\
\infty & \mbox{ if $\nu \npreceq \theta$.}
\end{cases}
\end{equation}
From the convexity of the $x \log x$ function for nonnegative $x$,
one can check that $D(\nu \| \theta) \ge 0$.

Here, and in the rest of the paper,
$:=$ is notation for equality by definition. 
Logarithms can be assumed to be to the natural base.
For two measurable functions $f$ and $g$ on $(S,\mF)$, not necessarily bounded, and 
$\eta \in \mP(S)$, $f =_\eta g$ denotes equality of $f$ and $g$ except possibly on an
$\eta$-null set. Similarly, for $C, D \in \mF$, $C =_\eta D$ denotes equality of $C$ and $D$
up to $\eta$-null sets and $C \subseteq_\eta D$ denotes the containment of $C$ in $D$ 
up to $\eta$-null sets.

The variational characterization in \eqref{eq:EasyVar} below of exponential integrals of bounded measurable
functions is elementary.
For any $\mu \in \mP(S)$ and 
$g \in \mB(S)$ we have
\begin{equation}		\mylabel{eq:EasyVar}
\log \int_S e^g d \mu = \sup_{\theta \in \mP(S)} \left( \int _S g d \theta - D(\theta \| \mu)
\right)
= \sup_{\theta \in \mP(S) ~:~\theta \preceq \mu} \left( \int _S g d \theta - D(\theta \| \mu)
\right)~.
\end{equation}
We provide a proof in Appendix \ref{app:EasyVarProof}.

For any $\alpha \in \mbR \backslash \{0,1\}$, and $\nu, \theta \in \mP(S)$,
the R\'{e}nyi divergence $R_\alpha(\nu \| \theta)$ is defined as in eqn. (2.1)
of \cite{ACD}, by first defining it for $\alpha > 0$, $\alpha \neq 1$, by 
\begin{equation}		\mylabel{eq:RenyiPos}
R_\alpha(\nu \| \theta) := \begin{cases} 
\infty & \mbox{ if $\alpha > 1$ and $\nu \npreceq \theta$}\\
\frac{1}{\alpha(\alpha -1)} \log \int_{\{\nu' \theta' > 0\}} (\frac{\nu'}{\theta'})^\alpha d \theta
& \mbox{ otherwise}~,
\end{cases}
\end{equation}
where $\nu' := \frac{d \nu}{d \eta}$ and $\theta' := \frac{d \theta}{d \eta}$, where
$\eta \in \mP(S)$ is an arbitrary probability distribution such that
$\nu \preceq \eta$ and $\theta \preceq \eta$. 
It is straightforward to check that every choice of $\eta$, subject to the absolute 
continuity conditions, results in the same value of the R\'{e}nyi entropy.
Then, for $\alpha < 0$, we use the 
definition
\begin{equation}		\mylabel{eq:RenyiNeg}
R_\alpha(\nu \| \theta) := R_{1 - \alpha}(\theta \| \nu)~.
\end{equation}

\begin{rem}
Even though the definition of $R_\alpha(\nu \| \theta)$ is broken up into cases above,
a single formula would work, if suitably interpreted. One could write
\[
R_\alpha(\nu \| \theta) = \frac{1}{\alpha(\alpha -1)} \log \int_S (\nu')^\alpha (\theta')^{1 -\alpha} d \eta~,~~\mbox{ for all $\alpha \in \mbR \backslash \{0, 1\}$.}
\]
In this formula, if $\eta( \nu' > 0, \theta' = 0) > 0$ and $\alpha > 1$, then because
$(\nu')^\alpha (\theta')^{1 - \alpha} = \frac{(\nu')^\alpha }{(\theta')^{\alpha -1}} = \infty$ on 
this event, we are forced to intepret $R_\alpha(\nu \| \theta)$ as being $\infty$.
A similar argument forces us to interpret $R_\alpha(\nu \| \theta)$ as $\infty$ if
$\eta( \nu' = 0, \theta' > 0) > 0$ and $\alpha < 0$. Rather than requiring of the reader
the mental gymnastics needed to keep track of such interpretations, we prefer to break
the discussion up into cases.
\end{rem}

\begin{rem}
It is clear that $R_\alpha(\nu \| \theta) \ge 0$ (possibly $\infty$) if $\alpha > 1$
or $\alpha < 0$. For $0 < \alpha <1$,
an application of H\"{o}lder's inequality with $p := \frac{1}{\alpha}$ and
$q := \frac{1}{1 -\alpha}$ (so $\frac{1}{p} + \frac{1}{q} =1$) gives
\begin{eqnarray*}
\int_{\{\nu' \theta' > 0\}} (\frac{\nu'}{\theta'})^\alpha d \theta
&=& \int_{\{\nu' \theta' > 0\}} (\nu')^\alpha (\theta')^{1-\alpha} d \eta\\
&=& \int_S (\nu')^\alpha (\theta')^{1-\alpha} d \eta\\
&\le& \left( \int_S \nu' d \eta \right)^\alpha \left( \int_S \theta' d \eta \right)^{1-\alpha}\\
&=& 1~.
\end{eqnarray*}
Hence we also have $R_\alpha(\nu \| \theta) \ge 0$ (possibly $\infty$) 
if $0 < \alpha < 1$. Note in particular that if $\eta( \nu' \theta' > 0) = 0$, then
$R_\alpha(\nu \| \theta) = \infty$ for all $\alpha \in \mbR \backslash \{0, 1\}$.
\end{rem}

\section{Statement of the main result in the i.i.d. case} \mylabel{sec:Statement}

Our main result in the i.i.d case is the following variational characterization of R\'{e}nyi divergence.

\begin{thm}		\mylabel{thm:Main}
Let $\alpha \in \mbR \backslash \{0,1\}$ and $\nu, \theta \in \mP(S)$. 
Then, 
if $\alpha > 1$,
we have
\begin{equation}		\mylabel{eq:MainBig}
R_\alpha(\nu \| \theta) = 
\sup_{ \{\mu \in \mP(S) ~:~ \mu \preceq \nu \} }
\left( \frac{1}{\alpha} D( \mu \| \theta) - \frac{1}{\alpha -1} D(\mu \| \nu) \right)~,
\end{equation}
while, if $0 < \alpha < 1$, we have
\begin{equation}		\mylabel{eq:MainSmall}
R_\alpha(\nu \| \theta) = 
\inf_{ \{\mu \in \mP(S) ~:~ \mu \preceq \nu, \mu \preceq \theta\} }
\left( \frac{1}{\alpha} D( \mu \| \theta) - \frac{1}{\alpha -1} D(\mu \| \nu) \right)~,
\end{equation}
and, 
if $\alpha < 0$,
we have
\begin{equation}		\mylabel{eq:MainNeg}
R_\alpha(\nu \| \theta) = 
\sup_{ \{\mu \in \mP(S) ~:~ \mu \preceq \theta \} }
\left( \frac{1}{\alpha} D( \mu \| \theta) - \frac{1}{\alpha -1} D(\mu \| \nu) \right)~.
\end{equation}
Further, when $0 < \alpha < 1$, one can find $\mu \in \mP(S)$, $\mu \preceq \nu$, 
$\mu \preceq \theta$, achieving 
the infimum on the RHS of \eqref{eq:MainSmall}, whenever 
$\{\mu \in \mP(S) ~:~ \mu \preceq \nu, \mu \preceq \theta\}$ is nonempty.
\hfill $\Box$
\end{thm}

\begin{rem}
The case by case structure of this result is partly 
a consequence of the normalization chosen
for the R\'{e}nyi divergences (which is necessary to make R\'{e}nyi divergence nonnegative) and
partly a consequence of the need to apply the correct absolute continuity conditions.
If it considered desirable to write a singe formula covering all cases, this can be 
done by considering $\Lambda_\alpha( \nu \| \theta) 
:= \alpha(\alpha -1) R_\alpha(\nu \| \theta)$, for $\alpha \in \mbR \backslash \{0,1\}$.
Then one has the single formula
\[
\Lambda_\alpha( \nu \| \theta) = 
\sup_{ \{\mu \in \mP(S) ~:~ \mu \preceq \nu \mbox{ or } \mu \preceq \theta \} }
\left( (\alpha -1) D( \mu \| \theta) - \alpha D(\mu \| \nu) \right)~,
\]
for all $\alpha \in \mbR \backslash \{0,1\}$. Note, however, that the set over which the 
supremum is being taken need not be convex in general. This is essential to avoid encountering
expressions of the form $\infty - \infty$.
\end{rem}


\section{Discussion} \mylabel{sec:Discussion}

Atar, Chowdhary and Dupuis \cite{ACD} have recently established a variational formula
for exponential integrals of bounded measurable functions.  This is established in 
two forms. For any $\alpha \in \mbR \backslash \{0,1\}$, $\nu \in \mP(S)$,
and $g \in \mB(S)$, eqn. (2.6)
of \cite{ACD} states that 
\begin{equation}		\mylabel{eq:ACDplus}
\frac{1}{\alpha -1} \log \int_S e^{(\alpha -1) g} d \nu
= \inf_{\theta \in \mP(S)} \left( \frac{1}{\alpha} \log \int_S e^{\alpha g} d \theta
+ R_\alpha( \nu \| \theta) \right)~,
\end{equation}
while eqn. (2.7) of \cite{ACD} states that for any $\alpha \in \mbR \backslash \{0,1\}$, $\theta \in \mP(S)$,
and $g \in \mB(S)$ we have
\begin{equation}		\mylabel{eq:ACDminus}
\frac{1}{\alpha} \log \int_S e^{\alpha g} d \theta
= \sup_{\nu \in \mP(S)} \left( \frac{1}{\alpha -1} \log \int_S e^{(\alpha -1) g} d \nu
- R_\alpha( \nu \| \theta) \right)~.
\end{equation}
It is straightforward to exhibit the equivalence of these two forms. For instance, 
assuming \eqref{eq:ACDplus}, let $\beta := 1 - \alpha$ and $h := - g$, and conclude that 
for all $\beta \in \mbR \backslash \{0,1\}$, $\nu \in \mP(S)$,
and $h \in \mB(S)$ we have
\[
- \frac{1}{\beta} \log \int_S e^{\beta h} d \nu
= \inf_{\theta \in \mP(S)} \left( \frac{1}{1 - \beta} \log \int_S e^{(\beta -1) h} d \theta
+ R_{1 - \beta} ( \nu \| \theta) \right)~,
\]
or equivalently that
\[
\frac{1}{\beta} \log \int_S e^{\beta h} d \nu 
= \sup_{\theta \in \mP(S)} \left( \frac{1}{\beta -1} \log \int_S e^{(\beta -1) h} d \theta
- R_\beta ( \theta \| \nu) \right)~,
\]
which is \eqref{eq:ACDminus}. One can similarly go in the opposite direction. 
We will therefore focus only on the form in \eqref{eq:ACDminus}.
As observed in Remark 2.3 of \cite{ACD}, taking the limit as $\alpha \to 1$
in \eqref{eq:ACDminus} recovers the elementary variational formula for 
exponential integrals of bounded measurable functions in \eqref{eq:EasyVar}.

The structure of Theorem \ref{thm:Main} is motivated by the variational characterization in \eqref{eq:ACDminus}. 
We will now demonstrate that Theorem \ref{thm:Main} is at least as strong as
\eqref{eq:ACDminus} 
by deriving \eqref{eq:ACDminus} from 
Theorem \ref{thm:Main} and the elementary variational formula \eqref{eq:EasyVar}.

First of all, we show that for any $\alpha \in \mbR \backslash \{0,1\}$, $\theta \in \mP(S)$,
and $g \in \mB(S)$ one can find $\nu \in \mP(S)$ achieving the supremum in 
\eqref{eq:ACDminus}. 
This proof does not depend on Theorem \ref{thm:Main} and \eqref{eq:EasyVar}.
In fact, the supremum is achieved by the choice $\frac{1}{Z} e^{-g} d \nu = d \theta$, where 
$Z$ is the normalization factor, and it is elementary to prove this. For completeness, a proof
is included in Appendix \ref{app:SupAchieved}.

It remains to prove that for any $\alpha \in \mbR \backslash \{0,1\}$, 
$g \in \mB(S)$, and $\theta, \nu \in \mP(S)$, we have
\begin{equation}		\mylabel{eq:ACDminusIneq}
\frac{1}{\alpha} \log \int_S e^{\alpha g} d \theta
\ge \frac{1}{\alpha -1} \log \int_S e^{(\alpha -1) g} d \nu
- R_\alpha( \nu \| \theta)~.
\end{equation}
Assuming the truth of Theorem \ref{thm:Main}, and using \eqref{eq:EasyVar}, 
this is proved in Appendix \ref{app:ACDminusIneq}.

\section{Proof of Theorem \ref{thm:Main}}		\mylabel{sec:Proof}

We now prove Theorem \ref{thm:Main}. 

Consider first the case $\alpha > 1$.
Suppose $\nu \npreceq \theta$. Then the LHS of \eqref{eq:MainBig} is 
$\infty$. Also, in this case, we can choose $\mu \in \mP(S)$ such that $\mu \preceq \nu$ but
$\mu \npreceq \theta$, which makes the RHS of \eqref{eq:MainBig} also equal to $\infty$. 
Thus we may assume that $\nu \preceq \theta$. Given $K > 0$ sufficiently large, define $\mu_K \in \mP(S)$ by
\[
\mu'_K := \frac{1}{Z_K} (\nu')^\alpha (\theta')^{1 - \alpha} 1( (\nu')^\alpha (\theta')^{1 - \alpha} \le K)
\]
where $\eta \in \mM(S \times S)$ is chosen such that $\theta \preceq \eta$, and we define
$\nu' := \frac{d \nu}{d \eta}$, $\theta' := \frac{d \theta}{d \eta}$, and $\mu'_K := \frac{d \mu_K}{d \eta}$. Further,
\[
Z_K := \int_{\{ (\nu')^\alpha (\theta')^{1 - \alpha} \le K\}} (\nu')^\alpha (\theta')^{1 - \alpha} d \eta~,
\]
and $K$ sufficiently large means that $Z_K > 0$.
We note that $\mu_K \preceq \nu$ (and so
$\mu_K \preceq \theta$).
Then
\begin{eqnarray*}
&&\frac{1}{\alpha} D( \mu_K \| \theta) - \frac{1}{\alpha -1} D(\mu_K \| \nu)\\
&&~~~~~= \frac{1}{\alpha} \int_{\{ \mu'_K > 0\}} \left( \log\frac{\mu'_K}{\theta'} \right) d \mu_K 
- \frac{1}{\alpha - 1} \int_{\{ \mu'_K > 0\}} \left( \log\frac{\mu'_K}{\nu'} \right) d \mu_K\\
&&~~~~~= \frac{1}{\alpha} \int_{\{ \mu'_K > 0\}} \left( \log\frac{(\nu')^\alpha}{Z_K (\theta')^\alpha} \right) d \mu_K 
- \frac{1}{\alpha -1} \int_{\{ \mu'_K > 0\}} \left( \log\frac{(\theta')^{1-\alpha}}{Z_K (\nu')^{1-\alpha}} \right) d \mu_K\\
&&~~~~~= \frac{1}{\alpha(\alpha -1)} \log Z_K~,
\end{eqnarray*}
which, as $K \to \infty$, converges to
\[
R_\alpha(\nu \| \theta) = \frac{1}{\alpha(\alpha -1)} \log \int_{\{\nu'\theta' > 0\}} \left( \frac{\nu'}{\theta'} \right)^\alpha d \theta = \frac{1}{\alpha(\alpha -1)} \log \int_S (\nu')^\alpha (\theta')^{1-\alpha} d \eta~.
\]
It remains to  show that, in the case $\alpha > 1$, 
for all $\nu, \theta \in \mP(S)$ such that $\nu \preceq \theta$, we have,
for all $\mu \in \mP(S)$ such that $\mu \preceq \nu$,
the inequality
\begin{equation}		\mylabel{eq:MainIneqBig}
R_\alpha(\nu \| \theta) \ge 
\frac{1}{\alpha} D( \mu \| \theta) - \frac{1}{\alpha -1} D(\mu \| \nu)~.
\end{equation}
Pick $\eta \in \mP(S)$ such that $\theta \preceq \eta$
(so we also have $\nu \preceq \eta$ and $\mu \preceq \eta$), and let $\nu' := \frac{d \nu}{d \eta}$, 
$\theta' := \frac{d \theta}{d \eta}$, and $\mu' := \frac{d \mu}{d \eta}$. Multiplying the RHS
of \eqref{eq:MainIneqBig} by $\alpha(\alpha -1)$ gives
\[
(\alpha -1) \int_{\{ \nu' \theta' \mu' > 0\}} \log \frac{\mu'}{\theta'}~~d \mu
- \alpha \int_{\{ \nu' \theta' \mu' > 0\}} \log \frac{\mu'}{\nu'}~~d \mu
= \int_{\{ \nu' \theta' \mu' > 0\}} \log \frac{(\nu')^\alpha (\theta')^{1 - \alpha}}{\mu'}~~ d \mu~.
\]
On the other hand, we have 
\[
\alpha (\alpha -1) R_\alpha(\nu \| \theta)  = 
\log \int_{\{\nu' \theta' > 0\}} (\frac{\nu'}{\theta'})^\alpha d \theta
\ge \log \int_{\{\nu' \theta' \mu' > 0\}} (\frac{\nu'}{\theta'})^\alpha \frac{\theta'}{\mu'} d \mu~,
\]
so \eqref{eq:MainIneqBig} follows from the concavity of the logarithm.

Next, consider the case when $0 < \alpha < 1$.
Pick $\eta \in \mP(S)$ such that $\nu \preceq \eta$ and $\theta \preceq \eta$, and let 
$\nu' := \frac{d \nu}{d \eta}$ and
$\theta' := \frac{d \theta}{d \eta}$.
If $\{ \nu' \theta' > 0\} =_\eta \emptyset$, then 
$\int_{\{\nu' \theta' > 0\}} (\frac{\nu'}{\theta'})^\alpha d \theta = 0$, and so 
\[
R_\alpha(\nu \| \theta) := 
\frac{1}{\alpha(\alpha -1)} \log \int_{\{\nu' \theta' > 0\}} (\frac{\nu'}{\theta'})^\alpha d \theta
= \infty~.
\]
But we also have
$\{ \mu \in \mP(S) ~:~ \mu \preceq \nu, \mu \preceq \theta \} = \emptyset$,
so the RHS of \eqref{eq:MainSmall} equals $\infty$.
We may therefore assume that $\eta( \nu' \theta' > 0) > 0$.
Now, an application of H\"{o}lder's inequality with $p := \frac{1}{\alpha}$ and
$q := \frac{1}{1 -\alpha}$ (so $\frac{1}{p} + \frac{1}{q} =1$) gives
\begin{eqnarray*}
\int_{\{\nu' \theta' > 0\}} (\frac{\nu'}{\theta'})^\alpha d \theta
&=& \int_{\{\nu' \theta' > 0\}} (\nu')^\alpha (\theta')^{1-\alpha} d \eta\\
&=& \int_S (\nu')^\alpha (\theta')^{1-\alpha} d \eta\\
&\le& \left( \int_S \nu' d \eta \right)^\alpha \left( \int_S \theta' d \eta \right)^{1-\alpha}\\
&=& 1
\end{eqnarray*}
Let $\mu \in \mP(S) $ be defined by $\mu' := \frac{1}{Z} (\nu')^\alpha (\theta')^{1-\alpha}$,
where $Z := \int_S (\nu')^\alpha (\theta')^{1-\alpha} d \eta$. Note that 
$R_\alpha(\nu \| \theta) = \frac{1}{\alpha (\alpha -1)} \log Z$.
We have $\mu \preceq \nu$ and $\mu \preceq \theta$, 
as required on the RHS of \eqref{eq:MainSmall}.
Now,
\begin{eqnarray*}
&&\frac{1}{\alpha} D( \mu \| \theta) - \frac{1}{\alpha -1} D(\mu \| \nu)\\
&&~~~~~= \frac{1}{\alpha} \int_{\{ \mu' > 0\}} \left( \log\frac{\mu'}{\theta'} \right) d \mu
- \frac{1}{\alpha - 1} \int_{\{ \mu' > 0\}} \left( \log\frac{\mu'}{\nu'} \right) d \mu\\
&&~~~~~= \frac{1}{\alpha} \int_{\{ \mu' > 0\}} \left( \log\frac{(\nu')^\alpha}{Z (\theta')^\alpha} \right) d \mu 
- \frac{1}{\alpha -1} \int_{\{ \mu' > 0\}} \left( \log\frac{(\theta')^{1-\alpha}}{Z (\nu')^{1-\alpha}} \right) d \mu\\
&&~~~~~= \frac{1}{\alpha(\alpha -1)} \log Z~,
\end{eqnarray*}
which equals $R_\alpha(\nu \| \theta)$.
It remains to  show that, in the case $0 < \alpha < 1$, 
for all $\nu, \theta \in \mP(S)$ such that $\eta( \nu' \theta' > 0) > 0$, we have,
for all $\mu \in \mP(S)$ such that $\mu \preceq \nu$ and $\mu \preceq \theta$,
the inequality
\begin{equation}		\mylabel{eq:MainIneqSmall}
R_\alpha(\nu \| \theta) \le 
\frac{1}{\alpha} D( \mu \| \theta) - \frac{1}{\alpha -1} D(\mu \| \nu)~.
\end{equation}
To see this, note that 
\begin{eqnarray*}
&& \alpha(1 - \alpha) \left( \frac{1}{\alpha} D( \mu \| \theta) - \frac{1}{\alpha -1} D(\mu \| \nu) \right)\\
&&~~~~~= (1 - \alpha) D( \mu \| \theta)  + \alpha D(\mu \| \nu)\\
&&~~~~~= (1 - \alpha) \int_{\{ \mu' > 0\}} \left( \log\frac{\mu'}{\theta'} \right) d \mu
+ \alpha \int_{\{ \mu' > 0\}} \left( \log\frac{\mu'}{\nu'} \right) d \mu\\
&&~~~~~= \int_{\{ \mu' > 0\}} \left( \log \frac{\mu'}{(\nu')^\alpha (\theta')^{1-\alpha}} \right) d \mu \\
&&~~~~~= \int_S f(\frac{(\nu')^\alpha (\theta')^{1-\alpha}}{\mu'}) \mu' d \eta\\
&&~~~~~\ge f \left( \int_S (\nu')^\alpha (\theta')^{1-\alpha} d \eta \right)\\
&&~~~~~= - \log \int_S (\nu')^\alpha (\theta')^{1-\alpha} d \eta \\
&&~~~~~= \alpha(1 - \alpha) R_\alpha( \nu \| \theta)~.
\end{eqnarray*}
where $f(\cdot)$ is the negative logarithm function, which is decreasing and convex.
This establishes \eqref{eq:MainIneqSmall}.
Note that we have also estabished the claim in Theorem \ref{thm:Main} that when $0 < \alpha < 1$
one can find $\mu$ realizing the infimum in \eqref{eq:MainSmall}
whenever 
$\{\mu \in \mP(S) ~:~ \mu \preceq \nu, \mu \preceq \theta\}$ is nonempty.

It remains to consider the case where $\alpha < 0$. Let $\beta := 1 - \alpha$. Then $\beta > 1$.
By definition $R_\alpha(\nu \| \theta) = R_\beta(\theta \| \nu)$. 
However,
we have already proved that 
\[
R_\beta( \theta \| \nu) = \sup_{\{\mu \in \mP(S) ~:~ \mu \preceq \theta\}}
\left( \frac{1}{\beta} D(\mu \| \nu) - \frac{1}{\beta -1} D( \mu \| \theta) \right)~.
\]
This reads
\[
R_\alpha(\nu \| \theta) = \sup_{\{\mu \in \mP(S) ~:~ \mu \preceq \theta\}}
\left( \frac{1}{\alpha} D( \mu \| \theta)  - \frac{1}{\alpha -1} D(\mu \| \nu) \right)~,
\]
which establishes \eqref{eq:MainNeg} in this case also and completes the proof of 
Theorem \ref{thm:Main}.

\section{R\'{e}nyi divergence rate between stationary finite state Markov chains}		\mylabel{sec:Markov}

In this section we set the stage to 
present analogs of the preceding results involving the R\'{e}nyi divergence
rates between two stationary finite state Markov chains. Extensions 
to general state space Markov processes in both discrete and continuous time of a form
similar to those we will present for stationary finite state Markov chains
no doubt exist, under
suitable conditions on the transition kernel, but may be considered topics
for future work.

From this point onwards in this document we take $S = \{1, \ldots, d\}$ and $\mF$ to be comprised of all the
subsets of $S$. Let $\mM(S \times S)$ denote the set of Markov probability distributions
on $(S \times S, \mF \times \mF)$, where $\nu \in \mM(S \times S)$ if $\nu(i,j) \ge 0$ for all $(i,j) \in S \times S$, $\sum_{i,j \in S} \nu(i,j) = 1$,
and 
$\nu(k, *)  = \nu( *, k)$ for all $k \in S$, where
$\nu(k, *) := \sum_{j \in S} \nu(k,j)$ and $\nu( *, k) := \sum_{i \in S} \nu(i, k)$.
Here $\mF \times \mF$ is comprised of all the subsets of $S \times S$.

Given $\nu \in \mM(S \times S)$, let $S_\nu := \{ k ~:~ \nu(k, *) > 0 \}$. 
$S_\nu$ is a subset of $S$, and is called the {\em support} of $\nu$. 
For $i \in S_\nu$ and
$j \in S$, we define $\nu(j|i) := \frac{\nu(i, j)}{\nu(i, \cdot)}$. Note that
$\nu(j|i) = 0$ if $i \in S_\nu$ and $j \notin S_\nu$, and $\sum_{j \in S} \nu(j|i) = 1$.
For $i \notin S_\nu$, we define $\nu(j|i) = 0$ for all $j$. This may seem strange, but
is an important notational convention for the equations we are going to write. Note that 
$\sum_{j \in S} \nu(j|i) = 0$ for $i \notin S_\nu$.

Given $\nu, \theta \in \mM(S \times S)$ we say $\nu$ is absolutely continuous with respect to 
$\theta$, denoted $\nu \preceq \theta$, if $\theta(i,j) = 0 \Rightarrow \nu(i,j) = 0$
for all $(i,j) \in S \times S$. 
The relative entropy $D(\nu \| \theta)$
of $\nu$ with respect to $\theta$ is defined by
\begin{equation}		\mylabel{eq:MarkovRelEnt}
D(\nu \| \theta) := \begin{cases} 
\sum_{i,j \in S_\nu} \nu(i,j) \log \frac{\nu(j|i)}{\theta(j|i)}~,& \mbox{ if $\nu \preceq \theta$,}\\
\infty & \mbox{ if $\nu \npreceq \theta$.}
\end{cases}
\end{equation}
It can be checked that $D(\nu \| \theta) \ge 0$.

We need certain basic facts about the asymptotic properties of iterated powers of
square matrices with nonnegative entries. We will state these facts in narrative form.
Proofs can be extracted from several books that provide 
standard treatments of the theory of nonnegative matrices or
finite state Markov chains, see e.g. \cite[Chap. 1]{Seneta}.

Let $M = \left[ m_{ij} \right]$ be a $d \times d$ matrix with nonnegative entries.
Then the limit
\begin{equation}		\mylabel{eq:GrowthRate}
\rho(M) := \lim_{n \to \infty} \frac{1}{n} \log \sum_{i,j} m^{(n)}(i,j)~,
\end{equation}
exists, where $m^{(n)}(i,j)$ denotes the $(i,j)$ entry of $M^n$.
We can associate to $M$ a directed graph on the vertex set $\{1, \dots, d\}$, where
we have a directed edge from $i$ to $j$ iff $m_{ij} > 0$. 
This graph may have self loops. Then $\rho(M) = - \infty$
iff this directed graph does not have a directed cycle. Otherwise $\rho(M)$ is finite. 
We call $\rho(M)$ the {\em growth rate} of $M$.

Suppose $\rho(M)$ is finite.
We say $\mu \in \mM(S \times S)$ is absolutely continuous with respect to $M$
if $\mu(i,j) > 0 \Rightarrow m(i,j) > 0$ for all $i,j \in S$
Let $\mu_1, \mu_2 \in \mM(S \times S)$ be absolutely continuous with respect to $M$.
Then so is $\frac{1}{2} (\mu_1 + \mu_2)$.
Thus there is a maximum element $\mu \in \mM(S \times S)$ that is absolutely continuous with 
respect to $M$, in the sense that every other $\nu \in \mM(S \times S)$ that is
absolutely continuous with respect to $M$ satisfies $\nu \preceq \mu$. 
This maximum element need
not be unique. Pick any such maximum element, call it $\tau$. 
Let $M' := \left[ m(i,j) 1(i, j \in S_\tau) \right]$.
Then $\rho(M') = \rho(M)$. 

Let $\mu \in \mM(S \times S)$, which we also think of as a nonnegative $d \times d$ matrix.
The support of $\mu$ can be uniquely written as a disjoint
union of subsets, called classes, 
$S_\mu = \cupdot_{k=1}^l C_k$, 
for some $l \ge 1$, such that 
$\mu(i,j) = 0$ if $i, j \in S_\mu$ are in distinct classes, and such that, for each $1 \le k \le l$,
if we consider the restriction of the directed graph associated to $\mu$
to the vertices in the class $C_k$, then this directed graph is 
irreducible, in the sense that there is a directed path in the graph between any pair of vertices
in $C_k$.

Given $\mu \in \mM(S \times S)$ and a $d \times d$ matrix $M$ with nonnegative entries,
we say $M$ is {\em compatible} with $\mu$ if $m(i,j) > 0 \Leftrightarrow \mu(i,j) > 0$. 
Let $S_\mu = \cupdot_{k=1}^l C_k$ be the decomposition of the support of $\mu$ into
classes. For each $1 \le k \le l$, the restriction of $M$ to the coordinates in $C_k$
defines a $|C_k| \times |C_k|$ irreducible matrix with nonnegative entries. 
This matrix has an associated Perron-Frobenius eigenvalue, which we denote by
$\lambda_k(M)$. We have $\lambda_k(M) > 0$ for all $1 \le k \le l$. 
We have $\rho(M) = \log \max_{1 \le k \le l} \lambda_k(M)$.
Also, for each
$1 \le k \le l$, the restriction of $M$ to 
the coordinates in $C_k$ has a left eigenvector associated
to the eigenvalue $\lambda_k(M)$, which has all its coordinates strictly positive
and is unique up to scaling, and also a 
right eigenvector associated
to the eigenvalue $\lambda_k(M)$, which has all its coordinates strictly positive
and is unique up to scaling.

Given $\nu \in \mM(S \times S)$, what we mean by the stationary Markov chain defined by 
$\nu$ is the following: for each $n \ge 1$ define a probability distribution
$\nu_n$ on $(S^n, \mF_n)$, where $\mF_n$ is comprised of all subsets of $S^n$, by setting
\begin{eqnarray*}
\nu_1(k) &=& \nu(k, *)~, \mbox{  for all $k \in S$}~,\\
\nu_2(i,j) &=& \nu(i,j)~, \mbox{  for all $i,j \in S$}~,\\
&\vdots& \\
\nu_n(i_1, \ldots, i_n) &=& \nu(i_1, i_2) \prod_{k=2}^{n-1} \nu(i_{k+1}|i_k)~, 
\mbox{ for all $(i_1, \ldots, i_n) \in S^n$}~,\\
&\vdots&~.\\
\end{eqnarray*}
It is straightfoward to check that for all $n \ge 2$ and $\nu, \theta \in \mM(S \times S)$
we have 
\begin{equation}		\mylabel{eq:AbsCont}
\nu \preceq \theta \Leftrightarrow \nu_n \preceq \theta_n~.
\end{equation}

The following fact, which will be very useful later, is easy to verify from the definitions.
It holds for all $\nu, \theta \in \mM(S \times S)$.
\begin{equation}		\mylabel{eq:RelEntLimit}
D( \nu \| \theta) = \lim_{n \to \infty} \frac{1}{n} D(\nu_n \| \theta_n)~,
\end{equation}
where on the RHS of this defintion the notation $D(\nu_n \| \theta_n)$ 
refers to the relative entropy between probability distributions on $(S^n, \mF_n)$.
 
We are now in a position where we can state the analog for stationary finite state Markov
chains of the elementary variational formula \eqref{eq:EasyVar}.
Let $G = \left[ g(i,j) \right] \in \mbR^{d \times d}$ and $\mu \in \mM(S \times S)$.
We have the following variational characterization of the growth rate
of the exponential integral of $G$ along the stationary Markov chain defined by $\mu$.

\begin{eqnarray}		\mylabel{eq:MarkovEasyVar}
\rho( \left[ e^{g(i,j)} \mu(j|i) \right] ) &=& \sup_{\theta \in \mM(S \times S)}
\left( \sum_{i,j \in S} g(i,j) \theta(i,j) - D( \theta \| \mu) \right) \\
&=& \sup_{\theta \in \mM(S \times S) ~:~ \theta \preceq \mu} 
\left( \sum_{i,j \in S} g(i,j) \theta(i,j) - D( \theta \| \mu) \right)~. \nonumber
\end{eqnarray}
The proof is in 
Appendix \ref{app:MarkovEasyVarProof}. The result is standard, being 
Varadhan's characterization of the spectral radius of nonnegative matrices, see e.g.
\cite[Exer. 3.1.19]{DZ}.

We are also in a position to define the R\'{e}nyi divergence rates between two stationary 
finite state Markov chains. This definition is classical, see e.g. the paper of Rached, Alajaji, and
Campbell \cite{RAC}, which also considers the nonstationary case, and the references therein.
Given $\nu, \theta \in \mM(S \times S)$ and $\alpha \in \mbR \backslash \{0,1\}$, we define
the R\'{e}nyi divergence rate of $\nu$ with respect to $\theta$, denoted $R_\alpha(\nu \| \theta)$,
by
\begin{equation}			\mylabel{eq:RenyiRateExists}
R_\alpha(\nu \| \theta) := \lim_{n \to \infty} \frac{1}{n} R_\alpha(\nu_n \| \theta_n)~,
\end{equation}
where on the RHS of this defintion the notation $R_\alpha(\nu_n \| \theta_n)$ 
refers to the R\'{e}nyi divergence between probability distributions on $(S^n, \mF_n)$
defined as in \eqref{eq:RenyiPos} and \eqref{eq:RenyiNeg}. The proofs of the
existence of the limit in 
\eqref{eq:RenyiRateExists} as well as of the  
properties of the R\'{e}nyi divergence rate of interest to us,
which are stated in the following proposition, are in Appendix \ref{app:RenyiRateProps}.

\begin{prop}		\mylabel{prop:RenyiRateProps}
Given $\nu, \theta \in \mM(S \times S)$, the R\'{e}nyi divergence rate, as defined in 
\eqref{eq:RenyiRateExists}, satisfies the following properties:
\[
R_\alpha(\nu \| \theta) = \begin{cases}
\infty & \mbox{ if $\alpha > 1$ and $\nu \npreceq \theta$}~,\\
\frac{1}{\alpha(\alpha -1)} \rho( \left[ \nu(j|i)^\alpha \theta(j|i)^{1 - \alpha} \right]) 
& \mbox{ if $0 < \alpha < 1$ or if $\alpha > 1$ and $\nu \preceq \theta$}~,
\end{cases}
\]
and 
\[
R_\alpha(\nu \| \theta) = R_{1 - \alpha} (\theta \| \nu)~,~~\mbox{ if $\alpha < 0$}~.
\]
\end{prop}

\section{Main results in the Markov case}		\mylabel{sec:MarkovMain}

Our first main result in the Markov case is the following variational characterization of the
R\'{e}nyi divergence rate, which is a direct analog of Theorem \ref{thm:Main}.

\begin{thm}		\mylabel{thm:MarkovMain}
Let $\alpha \in \mbR \backslash \{0,1\}$ and $\nu, \theta \in \mM(S \times S)$. 
Then, 
if $\alpha > 1$,
we have
\begin{equation}		\mylabel{eq:MarkovMainBig}
R_\alpha(\nu \| \theta) = 
\sup_{ \{\mu \in \mM(S \times S) ~:~ \mu \preceq \nu \} }
\left( \frac{1}{\alpha} D( \mu \| \theta) - \frac{1}{\alpha -1} D(\mu \| \nu) \right)~,
\end{equation}
while, if $0 < \alpha < 1$, we have
\begin{equation}		\mylabel{eq:MarkovMainSmall}
R_\alpha(\nu \| \theta) = 
\inf_{ \{\mu \in \mM(S \times S) ~:~ \mu \preceq \nu, \mu \preceq \theta\} }
\left( \frac{1}{\alpha} D( \mu \| \theta) - \frac{1}{\alpha -1} D(\mu \| \nu) \right)~,
\end{equation}
and, 
if $\alpha < 0$,
we have
\begin{equation}		\mylabel{eq:MarkovMainNeg}
R_\alpha(\nu \| \theta) = 
\sup_{ \{\mu \in \mM(S \times S) ~:~ \mu \preceq \theta \} }
\left( \frac{1}{\alpha} D( \mu \| \theta) - \frac{1}{\alpha -1} D(\mu \| \nu) \right)~.
\end{equation}
Further, one can find $\mu \in \mM(S \times S)$ achieving the extremum on the 
RHS in all three cases, except in the case where $0 < \alpha < 1$ and
$\{\mu \in \mM(S \times S) ~:~ \mu \preceq \nu, \mu \preceq \theta\}$ is empty.
\hfill $\Box$
\end{thm}

Our second main result in the Markov case is the following
analog of the variational formula of \cite{ACD}.

\begin{thm}		\mylabel{thm:MarkovACD}
For any $\alpha \in \mbR \backslash \{0,1\}$, $\nu \in \mM(S \times S)$,
and $G = \left[ g(i,j) \right] \in \mbR^{d \times d}$, 
we have
\begin{equation}		\mylabel{eq:MarkovACDplus}
\frac{1}{\alpha -1} \rho( \left[ e^{(\alpha -1) g(i,j)} \nu(j|i) \right] )
= \inf_{\theta \in \mM(S \times S)} \left( \frac{1}{\alpha} \rho( \left[ e^{\alpha g(i,j)} \theta(j|i) \right] )
+ R_\alpha( \nu \| \theta) \right)~,
\end{equation}
and for any $\alpha \in \mbR \backslash \{0,1\}$, $\theta \in \mM(S \times S)$,
and $G = \left[ g(i,j) \right] \in \mbR^{d \times d}$ we have
\begin{equation}		\mylabel{eq:MarkovACDminus}
\frac{1}{\alpha} \rho( \left[ e^{\alpha g(i,j)} \theta(j|i) \right] )
= \sup_{\nu \in \mM(S \times S)} \left( \frac{1}{\alpha -1} \rho( \left[ e^{(\alpha -1) g(i,j)} \nu(j|i) \right] )
- R_\alpha( \nu \| \theta) \right)~.
\end{equation}
\end{thm}

It is straightforward to exhibit the equivalence of the claims
in \eqref{eq:MarkovACDplus} and \eqref{eq:MarkovACDminus}. 
This is done is Appendix \ref{app:MarkovACDProofs}.
It therefore suffices to focus only on the form in \eqref{eq:MarkovACDminus}.
It is straightforward to show that for each $\theta \in \mM(S \times S)$ and
$G \in \mbR^{d \times d}$, one can find $\nu \in \mM(S \times S)$ achieving the supremum
on the RHS of \eqref{eq:MarkovACDminus}.
Appendix \ref{app:MarkovACDProofs} also contains a demonstration of this fact.
A proof of
Theorem \ref{thm:MarkovACD}, 
assuming the truth of Theorem \ref{thm:MarkovMain}, and using \eqref{eq:MarkovEasyVar}, is also provided in Appendix
\ref{app:MarkovACDProofs}.

\section{Proof of Theorem \ref{thm:MarkovMain}}		\mylabel{sec:MarkovProofMain}

Suppose $\alpha > 1$. If $\nu \npreceq \theta$, taking $\mu = \nu$ on the RHS of 
\eqref{eq:MarkovMainBig} makes the RHS equal $\infty$, which is also the value of the LHS.
We may therefore assume that $\nu \preceq \theta$.

Let $M := \left[ \nu(j|i)^\alpha \theta(j|i)^{1-\alpha} \right]$. This matrix is compatible
with $\nu$. 
Let $S_\nu = \cupdot_{k=1}^l C_k$ be the decomposition of the support of $\nu$ into
classes. We may choose the indexing of the classes in such a way that
$\rho(M) = \log \lambda_1(M)$.

Let $u$ be a $1 \times d$ row vector whose entries are zero in the coordinates that are not in 
$C_1$, while its restriction to $C_1$ is a nonzero left eigenvector of the restriction of 
$M$ to $C_1$. All the entries of $u$ in the coordinates in $C_1$ are strictly 
positive. Similarly, let $w$ be a $d \times 1$ 
column vector whose entries are zero in the coordinates that are not in 
$C_1$, while its restriction to $C_1$ is a nonzero right eigenvector of the restriction of 
$M$ to $C_1$. All the entries of $w$ in the coordinates in $C_1$ will be strictly 
positive.  For $i,j \in S$, we define
\[
\mu(i,j) := \frac{1}{Z} u(i) \nu(j|i)^\alpha \theta(j|i)^{1-\alpha}  w(j)~,
\]
where $Z := \sum_{i,j \in S} u(i) \nu(j|i)^\alpha \theta(j|i)^{1-\alpha}  w(j)$, 
which is strictly positive.
Note that  $\mu \in \mM(S \times S)$ and $\mu \preceq \nu$. We also have,
for all $i \in S$,
\[
\mu(i, *) := \sum_{j \in S} \mu(i,j) = \frac{1}{Z} \lambda_1(M) u(i) w(i)~,
\]
so we get
\[
\mu(j|i) = \begin{cases}
\frac{\nu(j|i)^\alpha \theta(j|i)^{1-\alpha} w(j)}{ \lambda_1(M) w(i)} & \mbox{ if $i,j \in C_1$}\\
0 & \mbox{ otherwise}~,
\end{cases}
\]
where we have used the fact that $S_\mu = C_1$.

Multiplying the RHS of \eqref{eq:MarkovMainBig} by $\alpha(\alpha -1)$
for this choice of $\mu$ gives
\begin{eqnarray*}
(\alpha -1) D(\mu \| \theta) - \alpha D(\mu \| \nu) 
&=&
\sum_{i,j \in C_1} \mu(i,j) \log \frac{\nu(j|i)^\alpha \theta(j|i)^{1-\alpha}}{\mu(j|i)}\\
&=& \sum_{i,j \in C_1} \mu(i,j) \log \frac{\lambda_1(M) w(i)}{w(j)}\\
&=& \log \lambda_1(M)~,
\end{eqnarray*}
which also equals $\alpha(\alpha -1)$ times the LHS of \eqref{eq:MarkovMainBig}.
This establishes the existence of $\mu \in \mM(S \times S)$ satisfying $\mu \preceq \nu$
and achieving equality in \eqref{eq:MarkovMainBig}.

It remains to check that for all $\mu \in \mM(S \times S)$ satisfying $\mu \preceq \nu$
we have the inequality 
\begin{equation}		\mylabel{eq:MarkovMainBigIneq}
R_\alpha(\nu \| \theta) \ge
\frac{1}{\alpha} D( \mu \| \theta) - \frac{1}{\alpha -1} D(\mu \| \nu) ~.
\end{equation}
But, in view of \eqref{eq:AbsCont}, in \eqref{eq:MainBig} applied to 
probability distributions on $(S^n, \mF_n)$, for $n \ge 2$, we have already proved 
that 
\[
R_\alpha(\nu_n \| \theta_n) \ge
\frac{1}{\alpha} D( \mu_n \| \theta_n) - \frac{1}{\alpha -1} D(\mu_n \| \nu_n) ~.
\]
Dividing by $n$, letting $n \to \infty$, and appealing to \eqref{eq:RelEntLimit} establishes \eqref{eq:MarkovMainBigIneq}.

Next, consider the case where $0 < \alpha < 1$. If the directed graph associated
to the matrix $M' := \left[ \nu(j|i)^\alpha \theta(j|i)^{1-\alpha} \right]$
has no cycles, then $R_\alpha(\nu \| \theta) = \infty$,  and
$\{ \mu \in \mM(S \times S) ~:~ \mu \preceq \nu, \mu \preceq \theta\} = \emptyset$,
so the RHS of \eqref{eq:MarkovMainSmall} is also $\infty$, and so \eqref{eq:MarkovMainSmall}
holds in this case. We may therefore assume that 
$\{ \mu \in \mM(S \times S) ~:~ \mu \preceq \nu, \mu \preceq \theta\}$ is nonempty.
Pick any $\tau \in \mM(S \times S)$ that is a maximum element among all the 
elements of $\mM(S \times S)$ that are absolutely continuous with respect to $M'$.
Let $M := \left[ \nu(j|i)^\alpha \theta(j|i)^{1-\alpha} 1(i, j \in S_\tau) \right]$.
Then $\rho(M') = \rho(M)$. Further, $M$ is compatible with $\tau$. 

Let $S_\tau = \cupdot_{k=1}^l C_k$ be the decomposition of the support of $\tau$ into
classes. We may choose the indexing of the classes in such a way that
$\rho(M) = \log \lambda_1(M)$.

Let $u$ be a $1 \times d$ row vector whose entries are zero in the coordinates that are not in 
$C_1$, while its restriction to $C_1$ is a nonzero left eigenvector of the restriction of 
$M$ to $C_1$. All the entries of $u$ in the coordinates in $C_1$ are strictly 
positive. Similarly, let $w$ be a $d \times 1$ 
column vector whose entries are zero in the coordinates that are not in 
$C_1$, while its restriction to $C_1$ is a nonzero right eigenvector of the restriction of 
$M$ to $C_1$. All the entries of $w$ in the coordinates in $C_1$ will be strictly 
positive.  For $i,j \in S$, we define
\[
\mu(i,j) := \frac{1}{Z} u(i) \nu(j|i)^\alpha \theta(j|i)^{1-\alpha}  w(j)~,
\]
where $Z := \sum_{i,j \in S} u(i) \nu(j|i)^\alpha \theta(j|i)^{1-\alpha}  w(j)$, 
which is strictly positive.
Note that  $\mu \in \mM(S \times S)$ and $\mu \preceq \tau$,
so $\mu \preceq \nu$ and $\mu \preceq \theta$. We also have,
for all $i \in S$,
\[
\mu(i, *) := \sum_{j \in S} \mu(i,j) = \frac{1}{Z} \lambda_1(M) u(i) w(i)~,
\]
so we get
\[
\mu(j|i) = \begin{cases}
\frac{\nu(j|i)^\alpha \theta(j|i)^{1-\alpha} w(j)}{ \lambda_1(M) w(i)} & \mbox{ if $i,j \in C_1$}\\
0 & \mbox{ otherwise}~,
\end{cases}
\]
where we have used the fact that $S_\tau = C_1$.

Multiplying the RHS of \eqref{eq:MarkovMainSmall} by $\alpha(1 - \alpha)$
for this choice of $\mu$ gives
\begin{eqnarray*}
(1 -\alpha) D(\mu \| \theta) + \alpha D(\mu \| \nu) 
&=&
\sum_{i,j \in C_1} \mu(i,j) \log \frac{\mu(j|i)}{\nu(j|i)^\alpha \theta(j|i)^{1-\alpha}}\\
&=& \sum_{i,j \in C_1} \mu(i,j) \log \frac{w(j)}{\lambda_1(M) w(i)}\\
&=& - \log \lambda_1(M)~,
\end{eqnarray*}
which also equals $\alpha(1- \alpha)$ times the LHS of \eqref{eq:MarkovMainSmall}.
This establishes the existence of $\mu \in \mM(S \times S)$ satisfying $\mu \preceq \nu$
and $\mu \preceq \theta$
and achieving equality in \eqref{eq:MarkovMainSmall}.

It remains to check that for all $\mu \in \mM(S \times S)$ satisfying $\mu \preceq \nu$
and $\mu \preceq \theta$
we have the inequality 
\begin{equation}		\mylabel{eq:MarkovMainSmallIneq}
R_\alpha(\nu \| \theta) \le
\frac{1}{\alpha} D( \mu \| \theta) - \frac{1}{\alpha -1} D(\mu \| \nu) ~.
\end{equation}
But, in view of \eqref{eq:AbsCont}, in \eqref{eq:MainSmall} applied to 
probability distributions on $(S^n, \mF_n)$, for $n \ge 2$, we have already proved 
that 
\[
R_\alpha(\nu_n \| \theta_n) \le
\frac{1}{\alpha} D( \mu_n \| \theta_n) - \frac{1}{\alpha -1} D(\mu_n \| \nu_n) ~.
\]
Dividing by $n$, letting $n \to \infty$, and appealing to \eqref{eq:RelEntLimit} establishes \eqref{eq:MarkovMainSmallIneq}.

It remains to consider the case $\alpha < 0$. Let $\beta := 1 - \alpha$. Then $\beta > 1$. 
By definition $R_\alpha(\nu \| \theta) = R_\beta(\theta \| \nu)$. 
However,
we have already proved that 
\[
R_\beta( \theta \| \nu) = \sup_{\{\mu \in \mP(S) ~:~ \mu \preceq \theta\}}
\left( \frac{1}{\beta} D(\mu \| \nu) - \frac{1}{\beta -1} D( \mu \| \theta) \right)~.
\]
This reads
\[
R_\alpha(\nu \| \theta) = \sup_{\{\mu \in \mP(S) ~:~ \mu \preceq \theta\}}
\left( \frac{1}{\alpha} D( \mu \| \theta)  - \frac{1}{\alpha -1} D(\mu \| \nu) \right)~,
\]
which establishes \eqref{eq:MarkovMainNeg} in this case also and completes the proof of 
Theorem \ref{thm:MarkovMain}.

\section{Concluding remarks}		\mylabel{sec:Conclusions}

We have given a variational characterization of R\'{e}nyi divergence 
between two arbitrary probability distributions on an arbitrary measurable space
in terms of relative entropies, for all values of the parameter defining the R\'{e}nyi divergence.
We also gave a variational characterization of the R\'{e}nyi divergence rate 
between 
two stationary finite state Markov chains
in terms of relative entropy rates, for all 
values of the parameter defining the R\'{e}nyi divergence rate.
A consequence of the latter development
was an analog of the variational formula of \cite{ACD} for stationary finite state Markov chains.

While we restricted ourselves
to stationary finite state Markov chains in the latter discussion,
it is to be expected that there will be
versions of this variational characterization of R\'{e}nyi divergence rate in a much 
broader setting involving Markov or $k$-th order Markov processes in discrete time,
and also in continuous time. 
It would also be interesting to consider to what extent such a 
variational characterization might generalize to the R\'{e}nyi divergence
rates between an arbitrary pair of stationary processes, assuming the existence of the
defining limit to start with,
since even the understanding of the relative entropy rate at this level of 
generality is somewhat limited \cite{Gray}.

\section*{Acknowledgments}

Thanks to Vivek Borkar and Payam Delgosha for their comments on a earlier draft of this
document.

\appendix

\section{Proof of the elementary variational formula in \eqref{eq:EasyVar}}	\mylabel{app:EasyVarProof}

The second equality in \eqref{eq:EasyVar} follows from the fact that $D(\theta \| \mu) = \infty$
if $\theta \npreceq \mu$.

Given $\mu \in \mP(S)$ and $g \in \mB(S)$, define $\theta \in \mP(S)$ by
$d \theta = \frac{1}{Z} e^g d \mu$, where $Z := \int_S e^g d \mu$. Note that $\theta \preceq \mu$.
Then
\[
\int_S g d \theta - D(\theta \| \mu) = \int_S g d \theta - \int_S \log \left(\frac{e^g}{Z}\right) d \theta = \log Z~,
\]
which also equals of the LHS of \eqref{eq:EasyVar}.

It remains to show that for all $\theta \preceq \mu$ we have 
\[
\log \int_S e^g d \mu \ge \int _S g d \theta - D(\theta \| \mu)~.
\]
Let $\theta' := \frac{d \theta}{d \mu}$.
We have
\[
\log \int_S e^g d \mu \ge \log \int_{\{\theta' > 0\}} \frac{e^g}{\theta'} d \theta
\ge \int_{\{\theta' > 0\}} \left( g - \log \theta' \right) d \theta 
= \int _S g d \theta - D(\theta \| \mu)~,
\]
where the second step is justified by the concavity of the logarithm.
This completes the proof.
\hfill $\Box$

\section{Proof that the supremum in \eqref{eq:ACDminus} is achieved}		\mylabel{app:SupAchieved}

Given $\theta \in \mP(S)$ and $g \in \mB(S)$, let $\nu \in \mP(S)$
be defined by $\frac{1}{Z} e^{-g} d \nu = d \theta$, where $Z := \frac{1}{\int e^g d \theta}$. 
Note that $\nu$ and $\theta$ are mutually absolutely continuous. 

Thus, for all 
$\alpha > 0$, $\alpha \neq 1$,
we have
\[
R_\alpha(\nu \| \theta) = \frac{1}{\alpha (\alpha -1)} \log \int_S Z^\alpha e^{\alpha g} d \theta
= \frac{1}{\alpha -1} \log Z + \frac{1}{\alpha (\alpha -1)} \log \int_S e^{\alpha g} d \theta~.
\]
On the other hand
\begin{eqnarray*}
\frac{1}{\alpha -1} \log \int_S e^{(\alpha -1) g} d \nu -
\frac{1}{\alpha} \log \int_S e^{\alpha g} d \theta
&=&\frac{1}{\alpha -1} \log \int_S Z e^{\alpha g} d \theta -
\frac{1}{\alpha} \log \int_S e^{\alpha g} d \theta\\
&=& \frac{1}{\alpha -1} \log Z + \frac{1}{\alpha (\alpha -1)} \log \int_S e^{\alpha g} d \theta~,
\end{eqnarray*}
which is the same.

Suppose now that $\alpha < 0$. Let $\beta := 1 - \alpha$. Then $\beta > 1$. For any
$\theta \in \mP(S)$ and $g \in \mB(S)$, let
$\nu \in \mP(S)$ be defined by $\frac{1}{Z} e^{-g} d \nu = d \theta$. Then 
$\frac{1}{W} e^{-h} d \theta = d \nu$, where $h := - g$ and 
$W =  \frac{1}{\int_S e^h d \nu} = \frac{1}{Z}$.
We have then already proved that
\begin{eqnarray*}
R_\alpha(\nu \| \theta) &:=& R_\beta(\theta \| \nu)\\
&=& \frac{1}{\beta -1} \log W + \frac{1}{\beta (\beta -1)} \log \int_S e^{\beta h} d \nu\\
&=& \frac{1}{\alpha} \log Z + \frac{1}{\alpha (\alpha -1)} \log \int_S e^{(\alpha -1) g} d \nu\\
&=& \frac{1}{\alpha -1} \log Z + \frac{1}{\alpha (\alpha -1)} \log \int_S e^{\alpha g} d \theta~,
\end{eqnarray*}
which completes the proof.
\hfill $\Box$

\section{Proof of \eqref{eq:ACDminusIneq}}		\mylabel{app:ACDminusIneq}

Consider first the case $\alpha > 1$. 
We may then assume that $\nu \preceq \theta$, since otherwise the right hand side 
of \eqref{eq:ACDminusIneq} is $- \infty$.
From \eqref{eq:EasyVar}, we have, for all 
$\mu \in \mP(S)$ such that $\mu \preceq \nu$ 
that
\[
\frac{1}{\alpha} \int_S e^{\alpha g} d \theta \ge \int_S g d \mu 
- \frac{1}{\alpha} D(\mu \| \theta)~.
\]
From \eqref{eq:MainBig} we have 
\[
R_\alpha(\nu \| \theta) \ge \frac{1}{\alpha} D( \mu \| \theta) - \frac{1}{\alpha -1} D(\mu \| \nu)~,
\]
which means that
\[
\frac{1}{\alpha} \int_S e^{\alpha g} d \theta \ge \int_S g d \mu - \frac{1}{\alpha -1} D(\mu \| \nu)
- R_\alpha(\nu \| \theta)~.
\]
Taking the supremum over $\mu \preceq \nu$ on the RHS of the preceding equation and 
using \eqref{eq:EasyVar} gives
\[
\frac{1}{\alpha} \int_S e^{\alpha g} d \theta \ge \frac{1}{\alpha -1} \log \int_S e^{(\alpha -1) g} d \nu
- R_\alpha( \nu \| \theta)~,
\]
which was to be shown.

Next, suppose $0 < \alpha < 1$. Given $g \in \mB(S)$ and $\nu, \theta \in \mP(S)$, 
if $\{ \nu' \theta' > 0 \} =_\eta \emptyset$ for some (and hence every) $\eta \in \mP(S)$
such that $\nu \preceq \eta$ and $\theta \preceq \eta$ (where $\nu' := \frac{d \nu}{d \eta}$
and $\theta' := \frac{d \theta}{d \eta}$), then $R_\alpha( \nu \| \theta) = \infty$,
and so \eqref{eq:ACDminusIneq} is true. Otherwise, we can find $\mu \in \mP(S)$
such that $\mu \preceq \nu$ and $\mu \preceq \theta$.
We know from the elementary variational formula \eqref{eq:EasyVar} that for
every $\mu \in \mP(S)$ we have
\[
\frac{1}{\alpha} \log \int_S e^{\alpha g} d \theta \ge \int_S g d \mu - \frac{1}{\alpha} D(\mu \| \theta)~,
\]
and
\[
\frac{1}{1 - \alpha} \log \int_S e^{(1 -\alpha) h} d \nu \ge \int_S h d \mu - \frac{1}{1 -\alpha} D(\mu \| \nu)~,
\]
where $h := -g$.
Hence
\[
\frac{1}{\alpha} \log \int_S e^{\alpha g} d \theta 
+ \frac{1}{1 - \alpha} \log \int_S e^{(1 -\alpha) h} d \nu 
\ge - \left( \frac{1}{\alpha} D(\mu \| \theta) -
\frac{1}{\alpha -1} D(\mu \| \nu) \right)~.
\]
But, from Theorem \ref{thm:Main}, we know that there exists $\mu \in \mP(S)$ for which
the RHS of the preceding equation equals $- R_\alpha(\nu \| \theta)$. This shows that
\[
\frac{1}{\alpha} \log \int_S e^{\alpha g} d \theta \ge
\frac{1}{\alpha -1} \log \int_S e^{(1 -\alpha) h} d \nu - R_\alpha( \nu \| \theta)~,
\]
which establishes \eqref{eq:ACDminusIneq} in this case.

It remains to consider the case $\alpha < 0$. Let $\beta := 1 - \alpha$, so $\beta > 1$.
We have already proved that 
\[
\frac{1}{\beta} \log \int_S e^{\beta h} d \nu \ge 
\frac{1}{\beta -1} \log \int_S e^{(\beta -1) h} d \theta - R_\beta(\theta \| \nu)~,
\]
where $h := - g$. Observing that $R_\beta(\theta \| \nu) = R_\alpha(\nu \| \theta)$,
this can be rewritten as
\[
\frac{1}{1 - \alpha} \log \int_S e^{(\alpha -1) g} d \nu 
\ge - \frac{1}{\alpha} \log \int_S e^{\alpha g} d \theta - R_\alpha(\nu \| \theta)~,
\]
which is \eqref{eq:ACDminusIneq} in this case, and completes the proof.
\hfill $\Box$

\section{Proof of \eqref{eq:MarkovEasyVar}}		\mylabel{app:MarkovEasyVarProof}

The second equality in \eqref{eq:MarkovEasyVar} follows from the fact that 
$D(\theta \| \mu) = \infty$
if $\theta \npreceq \mu$.

Given $\mu \in \mM(S \times S)$ and $G = \left[ g(i,j) \right] \in \mbR^{d \times d}$, 
the matrix $M :=\left[ e^{g(i,j)} \mu(j|i) \right]$ has nonnegative entries and is compatible 
with $\mu$, so $\rho(M)$, i.e.\ the LHS of \eqref{eq:MarkovEasyVar}, is finite.
Let $S_\mu = \cupdot_{k=1}^l C_k$ be the decomposition of the support of $\mu$ into
classes. We may choose the indexing of the classes in such a way that
$\rho(M) = \log \lambda_1(M)$.

Let $u$ be a $1 \times d$ row vector whose entries are zero in the coordinates that are not in 
$C_1$, while its restriction to $C_1$ is a nonzero left eigenvector of the restriction of 
$M$ to $C_1$. Note that all the entries of $u$ in the coordinates in $C_1$ are strictly 
positive. Similarly, let $w$ be a $d \times 1$ 
column vector whose entries are zero in the coordinates that are not in 
$C_1$, while its restriction to $C_1$ is a nonzero right eigenvector of the restriction of 
$M$ to $C_1$. All the entries of $w$ in the coordinates in $C_1$ will be strictly 
positive.  For $i,j \in S$, we define
\[
\theta(i,j) := \frac{1}{Z} u(i) e^{g(i,j)} \mu(j|i) w(j)~,
\]
where $Z := \sum_{i,j \in S} u(i) e^{g(i,j)} \mu(j|i) w(j)$, which is strictly positive.
Note that  $\theta \in \mM(S \times S)$ and $\theta \preceq \mu$. We also have,
for all $i \in S$,
\[
\theta(i, *) := \sum_{j \in S} \theta(i,j) = \frac{1}{Z} \lambda_1(M) u(i) w(i)~,
\]
so we get
\[
\theta(j|i) = \begin{cases}
\frac{e^{g(i,j)} \mu(j|i) w(j)}{ \lambda_1(M) w(i)} & \mbox{ if $i,j \in C_1$}\\
0 & \mbox{ otherwise}~,
\end{cases}
\]
where we have used the fact that $S_\theta = C_1$.

We may now compute
\begin{eqnarray*}
\sum_{i,j \in S} g(i,j) \theta(i,j)  - D(\theta \| \mu) &=& \sum_{i,j \in S} g(i,j) \theta(i,j) 
- \sum_{i,j \in C_1} \theta(i,j) \log \left(\frac{e^{g(i,j)} w(j)}{ \lambda_1(M) w(i)}\right)\\
&=& \sum_{i,j \in C_1} w(i) \theta(i,j) - \sum_{i,j \in C_1} \theta(i,j) w(j) + \log \lambda_1(M)\\
&=& \rho(M)~,
\end{eqnarray*}
which also equals of the LHS of \eqref{eq:MarkovEasyVar}. 
This establishes that for each
$\mu \in \mM(S \times S)$ and $G = \left[ g(i,j) \right] \in \mbR^{d \times d}$ 
there exists $\theta \in \mM(S \times S)$ achieving equality in \eqref{eq:MarkovEasyVar}.

It remains to show that for all $\theta \in \mM(S \times S)$ such that 
$\theta \preceq \mu$ we have 
\begin{equation}		\mylabel{eq:Varadhan}
\rho( \left[ e^{g(i,j)} \mu(j|i) \right] ) \ge \sum_{i,j \in S} g(i,j) \theta(i,j) - D(\theta \| \mu)~.
\end{equation}
But, 
using \eqref{eq:EasyVar} applied to the 
probability distribution $\mu_n$ on $(S^n, \mF_n)$, for $n \ge 2$, with 
$g(i_1, \ldots, i_n) := \sum_{k=1}^{n-1} g(i_k, i_{k+1})$, we have already proved that
\[
\log \sum_{i_1, \ldots, i_n} \mu(i_1, *) \prod_{k=1}^{n-1} e^{g(i_k, i_{k+1})} \mu(i_{k+1}|i_k)
\ge \sum_{i_1, \ldots, i_n} \sum_{k=1}^{n-1} g(i_k, i_{k+1}) \mu_n(i_1, \ldots, i_n) 
- D(\theta_n \| \mu_n)~.
\]
Divide both sides by $n$ and take the limit as $n \to \infty$.
Appealing to \eqref{eq:RelEntLimit} and the definition of the growth rate in
\eqref{eq:GrowthRate} proves 
\eqref{eq:Varadhan}. This completes the proof of \eqref{eq:MarkovEasyVar}.
\hfill $\Box$

\section{Proof of the existence of the limit in \eqref{eq:RenyiRateExists}, and of Proposition \ref{prop:RenyiRateProps}}		\mylabel{app:RenyiRateProps}

Suppose $\alpha > 1$ and $\nu \npreceq \theta$. Then $\nu_n \npreceq \theta_n$
for all $n \ge 2$ and so the limit on the RHS of \eqref{eq:RenyiRateExists} exists
and equals $\infty$, as claimed in  Proposition \ref{prop:RenyiRateProps}.

If $\alpha > 1$ and $\nu \preceq \theta$, then $\nu_n \preceq \theta_n$ for all 
$n \ge 2$, and so
\[
R_\alpha(\nu_n \| \theta_n) = \frac{1}{\alpha(\alpha -1)} 
\sum_{i_1, \ldots, i_n} \left( \nu(i_1,i_2) \prod_{k=2}^{n-1} \nu(i_{k+1}|i_k) \right)^\alpha
\left( \theta(i_1,i_2) \prod_{k=2}^{n-1} \theta(i_{k+1}|i_k) \right)^{1 -\alpha}~.
\]
This is also the formula for $R_\alpha(\nu_n \| \theta_n)$ when $0 < \alpha < 1$, 
irrespective of whether $\nu \preceq \theta$ or not. It follows from the definition of
the growth rate in \eqref{eq:GrowthRate} that the limit 
on the RHS of \eqref{eq:RenyiRateExists} exists
and equals $\frac{1}{\alpha(\alpha -1)} \rho( \left[ \nu(j|i)^\alpha \theta(j|i)^{1-\alpha} \right] )$, as claimed in Proposition \ref{prop:RenyiRateProps}.

Finally, suppose $\alpha < 0$. Let $\beta := 1 - \alpha$. Then we have $\beta > 1$.
We have therefore already proved that 
$\lim_{n \to \infty} \frac{1}{n} R_\beta( \theta_n \| \nu_n)$ exists and equals
$R_{1 - \alpha}(\theta \| \nu)$, as given in Proposition \ref{prop:RenyiRateProps}. But 
$R_\beta( \theta_n \| \nu_n)$ equals $R_\alpha(\nu_n \| \theta_n)$. Therefore the 
limit on the RHS of \eqref{eq:RenyiRateExists} exists, and since this is what we call
$R_\alpha(\nu \| \theta)$ it must be the case that $R_\alpha(\nu \| \theta)$ equals
$R_{1-\alpha}(\theta \| \nu)$, as claimed in 
Proposition \ref{prop:RenyiRateProps}. This completes the proof.
\hfill  $\Box$

\section{Proof of Theorem \ref{thm:MarkovACD} assuming the truth of Theorem \ref{thm:MarkovMain} and using \eqref{eq:MarkovEasyVar}, and proofs of the two claims about
\eqref{eq:MarkovACDminus} }		\mylabel{app:MarkovACDProofs}

We first verify the truth of the two claims about \eqref{eq:MarkovACDminus} 
which were made just after the statement of Theorem \ref{thm:MarkovACD}.

To exhibit the equivalence of the two forms \eqref{eq:MarkovACDplus}
and \eqref{eq:MarkovACDminus} appearing in Theorem  \ref{thm:MarkovACD}, assume,
for instance, the truth of \eqref{eq:MarkovACDplus}. Let $\beta := 1 - \alpha$ and $H 
= \left[ h(i,j) \right] = - G$, and conclude that 
for all $\beta \in \mbR \backslash \{0,1\}$, $\nu \in \mM(S \times S)$,
and $H \in \mbR^{d \times d}$ we have
\[
- \frac{1}{\beta} \rho( \left[ e^{\beta h(i,j)} \nu(j|i) \right] )
= \inf_{\theta \in \mM(S \times S)} \left( \frac{1}{1 - \beta} \rho( \left[ e^{(\beta -1) h(i,j)} \theta(j|i) \right] )
+ R_{1 - \beta} ( \nu \| \theta) \right)~,
\]
or equivalently that
\[
\frac{1}{\beta} \rho( \left[ e^{\beta h(i,j)} \nu(j|i)  \right] )
= \sup_{\theta \in \mM(S \times S)} \left( \frac{1}{\beta -1} \rho( \left[ e^{(\beta -1) h(i,j)} \theta(j|i) \right] )
- R_\beta ( \theta \| \nu) \right)~,
\]
which is \eqref{eq:MarkovACDminus}. One can similarly go in the opposite direction.

To verify that the supremum on the RHS of \eqref{eq:MarkovACDminus} is achieved, 
given $\theta \in \mM(S \times S)$, $G = \left[ g(i,j) \right] \in \mbR^{d \times d}$,
and $\alpha \in \mbR \backslash \{0,1\}$,
observe that $N := \left[ e^{\alpha g(i,j)} \theta (j|i)  \right]$ is compatible
with $\theta$.
Let $S_\mu = \cupdot_{k=1}^l C_k$ be the decomposition of the support of $\theta$ into
classes. We may choose the indexing of the classes in such a way that
$\rho(N) = \log \lambda_1(N)$.

Let $M := \left[ e^{g(i,j)} \theta (j|i)  \right]$.
Observe that $M$ is also compatible with $\theta$.
Let $u$ be a $1 \times d$ row vector whose entries are zero in the coordinates that are not in 
$C_1$, while its restriction to $C_1$ is a nonzero left eigenvector of the restriction of 
$M$ to $C_1$. All the entries of $u$ in the coordinates in $C_1$ are strictly 
positive. Similarly, let $w$ be a $d \times 1$ 
column vector whose entries are zero in the coordinates that are not in 
$C_1$, while its restriction to $C_1$ is a nonzero right eigenvector of the restriction of 
$M$ to $C_1$. All the entries of $w$ in the coordinates in $C_1$ will be strictly 
positive.  For $i,j \in S$, we define
\[
\nu(i,j) := \frac{1}{Z} u(i) e^{g(i,j)} \mu(j|i) w(j)~,
\]
where $Z := \sum_{i,j \in S} u(i) e^{g(i,j)} \mu(j|i) w(j)$, which is strictly positive.
Note that $\nu \in \mM(S \times S)$ and $\nu \preceq \theta$. 
We also have,
for all $i \in S$,
\[
\nu(i, *) := \sum_{j \in S} \nu(i,j) = \frac{1}{Z} \lambda_1(M) u(i) w(i)~,
\]
so we get
\[
\nu(j|i) = \begin{cases}
\frac{e^{g(i,j)} \theta(j|i) w(j)}{ \lambda_1(M) w(i)} & \mbox{ if $i,j \in C_1$}\\
0 & \mbox{ otherwise}~,
\end{cases}
\]
where we have used the fact that $S_\nu = C_1$.

We now note that
\[
\nu(j|i)^\alpha \theta(j|i)^{1 - \alpha} = \begin{cases}
\frac{e^{\alpha g(i,j)} \theta(j|i) w(j)^\alpha}{ \lambda_1(M)^\alpha w(i)^\alpha} & \mbox{ if $i,j \in C_1$}\\
0 & \mbox{ otherwise}~.
\end{cases}
\]
Then we have 
\begin{eqnarray}		\mylabel{eq:Rhos}
\rho( \left[ \nu(j|i)^\alpha \theta(j|i)^{1 - \alpha} \right] ) 
&=& \rho( \left[ e^{\alpha g(i,j)} \theta(j|i) 1(i, j \in C_1) \right] ) - \alpha \rho(M)~,
\nonumber \\
&=& \rho( \left[ e^{\alpha g(i,j)} \theta(j|i) \right] - \alpha \rho(M)~, \nonumber \\
&=& \rho(N) - \alpha \rho(M)
\end{eqnarray}
Here the first step can be seen by observing that the $w(i)^\alpha$ terms for $i \in C_1$
cancel each other out by successive cancellation in the defintion of the growth rate as a limit.
Equality in the second step depends on the fact that we have chosen $C_1$ such that
$\rho(N) = \log \lambda_1(N)$.

We also note that
\[
e^{(\alpha -1)g(i.j)} \nu(j|i) = \begin{cases}
\frac{e^{\alpha g(i,j)} \theta(j|i) w(j)}{ \lambda_1(M) w(i)} & \mbox{ if $i,j \in C_1$}\\
0 & \mbox{ otherwise}~,
\end{cases}
\]
so we have 
\begin{eqnarray}		\mylabel{eq:Rhos2}
\rho( \left[ e^{(\alpha -1)g(i.j)} \nu(j|i) \right] ) &=&
\rho( \left[ \frac{e^{\alpha g(i,j)} \theta(j|i) w(j)}{ \lambda_1(M) w(i)} 1(i, j \in C_1)
\right] )~, \nonumber \\
&=& \rho( \left[ e^{\alpha g(i,j)} \theta(j|i) 1(i, j \in C_1)
\right] ) - \rho(M)~, \nonumber \\
&=& \rho( \left[ e^{\alpha g(i,j)} \theta(j|i) \right] ) - \rho(M)~, \nonumber \\
&=& \rho(N) - \rho(M)~.
\end{eqnarray}
Here the first step can be seen by observing that the $w(i)$ terms for $i \in C_1$
cancel each other out by successive cancellation in the defintion of the growth rate as a limit,
and
equality in the second step depends on the fact that we have chosen $C_1$ such that
$\rho(N) = \log \lambda_1(N)$.

Since $\nu \preceq \theta$, we have
\[
R_\alpha( \nu \| \theta) = \frac{1}{\alpha(\alpha -1)} \rho( \left[ \nu(j|i)^\alpha \theta(j|i)^{1 - \alpha} \right] )~.
\]
Multiplying \eqref{eq:Rhos} through by $\frac{1}{\alpha(\alpha -1)}$ and
using \eqref{eq:Rhos2} gives
\[
R_\alpha( \nu \| \theta) = \frac{1}{\alpha -1} \rho( \left[ e^{(\alpha -1)g(i.j)} \nu(j|i) \right] )
- \frac{1}{\alpha} \rho(N)~,
\]
which demonstrates that $\nu$ works to show what what was claimed.

In order to prove Theorem \ref{thm:MarkovACD}, it remains to show that for 
every $\theta, \nu \in \mM(S \times S)$, $G = \left[ g(i,j) \right] \in \mbR^{d \times d}$,
and $\alpha \in \mbR \backslash \{0,1\}$, we have 
\begin{equation}		\mylabel{eq:MarkovACDminusIneq}
\frac{1}{\alpha} \rho( \left[ e^{\alpha g(i,j)} \theta(j|i) \right] )
\ge  \frac{1}{\alpha -1} \rho( \left[ e^{(\alpha -1) g(i,j)} \nu(j|i) \right] )
- R_\alpha( \nu \| \theta) ~.
\end{equation}
We prove this, 
assuming the truth of Theorem \ref{thm:MarkovMain}, using \eqref{eq:MarkovEasyVar}.
The proof is almost a verbatim copy of that in Appendix \ref{app:ACDminusIneq}, except
that we are now dealing with the case of stationary finite state Markov chains rather than
with the i.i.d. case.

Consider first the case $\alpha > 1$. 
We may then assume that $\nu \preceq \theta$, since otherwise the right hand side 
of \eqref{eq:MarkovACDminusIneq} is $- \infty$.
From \eqref{eq:MarkovEasyVar}, we have, for all 
$\mu \in \mM(S \times S)$ such that $\mu \preceq \nu$ 
that
\[
\frac{1}{\alpha} \rho( \left[ e^{\alpha g(i,j)} \theta(j|i) \right] ) \ge 
\sum_{i,j \in S} g(i,j) \mu(i,j)
- \frac{1}{\alpha} D(\mu \| \theta)~.
\]
From \eqref{eq:MarkovMainBig} we have 
\[
R_\alpha(\nu \| \theta) \ge \frac{1}{\alpha} D( \mu \| \theta) - \frac{1}{\alpha -1} D(\mu \| \nu)~,
\]
which means that
\[
\frac{1}{\alpha} \rho( \left[ e^{\alpha g(i,j)} \theta(j|i) \right] )  \ge \sum_{i,j \in S} g(i,j) \mu(i,j) - \frac{1}{\alpha -1} D(\mu \| \nu)
- R_\alpha(\nu \| \theta)~.
\]
Taking the supremum over $\mu \preceq \nu$ on the RHS of the preceding equation and 
using \eqref{eq:MarkovEasyVar} gives
\[
\frac{1}{\alpha} \rho( \left[ e^{\alpha g(i,j)} \theta(j|i) \right] ) \ge \frac{1}{\alpha -1} \rho( \left[ e^{(\alpha -1) g(i,j)} \nu(j|i) \right] )
- R_\alpha( \nu \| \theta)~,
\]
which was to be shown.

Next, suppose $0 < \alpha < 1$. There is no $\mu \in \mM(S \times S)$ such that 
$\mu \preceq \nu$ and $\mu \preceq \theta$ precisely when the directed graph
associated to $\left[ \nu(i,j)^\alpha \theta(i,j)^{1 - \alpha} \right]$ has no 
cycles, and in this case $R_\alpha( \nu \| \theta) = \infty$, so \eqref{eq:MarkovACDminusIneq} is true. Therefore, we may assume that we can find $\mu \in \mP(S)$
such that $\mu \preceq \nu$ and $\mu \preceq \theta$.
We know from \eqref{eq:MarkovEasyVar} that for
every $\mu \in \mM(S \times S)$ we have
\[
\frac{1}{\alpha} \rho( \left[ e^{\alpha g(i,j)} \theta(j|i) \right] ) \ge \sum_{i,j \in S} g(i,j) \mu(i,j) - \frac{1}{\alpha} D(\mu \| \theta)~,
\]
and
\[
\frac{1}{1 - \alpha} \rho( \left[ e^{(1 -\alpha) h(i,j)} \nu(j|i) \right] ) \ge \sum_{i,j \in S} h(i,j) \mu(i,j) - \frac{1}{1 -\alpha} D(\mu \| \nu)~,
\]
where $h := -g$.
Hence
\[
\frac{1}{\alpha} \rho( \left[ e^{\alpha g(i,j)} \theta(j|i) \right] ) 
+ \frac{1}{1 - \alpha} \rho( \left[ e^{(1 -\alpha) h(i,j)} \nu(j|i) \right] ) 
\ge - \left( \frac{1}{\alpha} D(\mu \| \theta) -
\frac{1}{\alpha -1} D(\mu \| \nu) \right)~.
\]
But, from Theorem \ref{thm:MarkovMain}, we know that there exists $\mu \in \mM(S \times S)$ for which
the RHS of the preceding equation equals $- R_\alpha(\nu \| \theta)$. This shows that
\[
\frac{1}{\alpha} \rho( \left[ e^{\alpha g(i,j)} \theta(j|i) \right] )  \ge
\frac{1}{\alpha -1} \rho( \left[ e^{(1 -\alpha) h(i,j)} \nu(j|i) \right] ) - R_\alpha( \nu \| \theta)~,
\]
which establishes \eqref{eq:MarkovACDminusIneq} in this case.

It remains to consider the case $\alpha < 0$. Let $\beta := 1 - \alpha$, so $\beta > 1$.
We have already proved that 
\[
\frac{1}{\beta} \rho( \left[ e^{\beta h(i,j)} \nu(j|i) \right] ) \ge 
\frac{1}{\beta -1} \rho( \left[ e^{(\beta -1) h(i,j)} \theta(j|i) \right] ) - R_\beta(\theta \| \nu)~,
\]
where $h := - g$. Observing that $R_\beta(\theta \| \nu) = R_\alpha(\nu \| \theta)$,
this can be rewritten as
\[
\frac{1}{1 - \alpha} \rho( \left[ e^{(\alpha -1) g(i,j)} \nu(j|i) \right] ) 
\ge - \frac{1}{\alpha} \rho( \left[ e^{\alpha g(i,j)} \theta(j|i) \right] ) - R_\alpha(\nu \| \theta)~,
\]
which is \eqref{eq:MarkovACDminusIneq} in this case, and completes the proof.
\hfill $\Box$

\end{document}